\documentclass[aps,pra,superscriptaddress,tightenlines]{revtex4}
\usepackage{graphicx} 
\usepackage{amsmath}
 
\newcommand{\beq}{\begin{equation}}
\newcommand{\enq}{\end{equation}}

\begin{document}

\title{Excitations of a Bose-Einstein condensate 
in a one-dimensional optical lattice}
\author{J.-P. Martikainen}
\email{J.P.J.Martikainen@phys.uu.nl}
\author{H. T. C. Stoof}
\email{stoof@phys.uu.nl}
\affiliation{Institute for Theoretical Physics, Utrecht University, 
Leuvenlaan 4, 3584 CE Utrecht, The Netherlands}
\date{\today}

\begin{abstract}
We investigate the low-lying excitations of a stack of weakly coupled
two-dimensional Bose-Einstein condensates,
which is formed by 
a one-dimensional optical lattice. In particular, we calculate
the dispersion relations of the monopole and quadrupole modes, 
both for the ground state as well as for the case
in which the system contains a vortex along the direction of the lasers
creating the optical lattice. 
Our variational approach enables us to determine analytically 
the dispersion relations
for an arbitrary number of atoms in every two-dimensional condensate 
and for an arbitrary momentum.
We also discuss the feasibility of observing
our results experimentally.
\end{abstract}
%\narrowtext
\maketitle

\section{Introduction}
\label{sec:intro}
Cold bosons in an optical lattice provide a uniquely tunable environment
to explore quantum phenomena. Some of these phenomena have been known 
theoretically
for quite some time, but with the advent of new experimental
tools they have become a focus of attention. 
For example, Bloch oscillations of electrons in a metal 
are standard material in
condensed-matter text books, but advances in the manipulation
of cold atoms have made their experimental investigation 
also possible in this case~\cite{Cristiani2002a}. 
In addition, diffraction of matter waves by  a pulsed optical lattice
was studied by Ovchinnikov {\it et al.}~\cite{Ovchinnikov1999a}. 
%~\cite{Kozuma1999a,Ovchinnikov1999a,StamperKurn1999a,Simsarian2000a,Ozeri2002a}.
Optical lattices have also enabled the observation of some 
more exotic quantum phenomena such as 
number squeezing~\cite{Orzel2001a} and collapses and 
revivals~\cite{Greiner2002b}. Apart from these examples, 
Bose-Einstein condensates in optical lattices are particularly promising
physical systems to study superfluid properties of Bose 
gases~\cite{Cataliotti2001a,Burger2001a}. Very importantly 
they realize the Bose-Hubbard model~\cite{Jaksch1998a} and 
can be used to investigate the quantum
phase transition from the superfluid into a 
Mott-insulator state~\cite{Jaksch1998a,vanOosten2001a}.
This phase transition was recently indeed 
observed experimentally~\cite{Greiner2002a}.

In this paper we present a variational method to study the excitations in 
a stack of weakly-coupled two-dimensional (quasi)condensates. 
Such a system can be created
by applying a relatively strong one-dimensional
optical lattice to an ordinary three-dimensional condensate.
We focus on the transverse monopole and quadrupole-modes, 
but we also demonstrate how
the method can be applied to study longitudinal excitations.
We determine the eigenfrequencies of the 
monopole and quadrupole modes without any other approximations than
those involved in our variational ansatz. 
In particular,
this means that we can smoothly crossover from the noninteracting limit
to the Thomas-Fermi regime. Moreover, the longitudinal 
wave length of the modulation
is arbitrary, i.e., nearest neighbor sites can be completely out of phase.
Using typical experimental parameters, 
we predict that the dispersion relations show a strong dependence on the
lattice potential. 

We consider the transverse excitations both for the ground state of the gas,
as well as for the case that a vortex pierces through the center of
each two-dimensional (quasi)condensate. We consider also the latter
case, because a recent experiment observed the 
transverse vibrational modes of the vortex line in a trapped 
Bose-Einstein condensate. 
These modes are called Kelvin modes~\cite{Bretin2003a}.
In the lattice three-dimensional 
effects, such as vortex line curvature, are expected
to be less important. Therefore, 
we believe that Kelvin modes, and 
in particular their coupling to the transverse
collective modes, are easier to study in a lattice 
than in a cigar-shaped three-dimensional condensate. Our results
for a stationary vortex represent the first necessary step 
toward understanding the more complicated problem of the Kelvin modes
of a vortex in an optical lattice.

There is a number of 
important theoretical papers on the dynamics of Bose-Einstein condensates 
in optical lattices. For example, dynamical and modulational instabilities
were studied in Refs.~\cite{Wu2001a,Konotop2002a,Baizakov2002a} and 
the adiabaticity of the nonlinear wave equations was 
explored by Band and Trippenbach~\cite{Band2002a}. 
Massignan and Modugno derived~\cite{Massignan2003a} 
a relatively simple way to solve the three-dimensional 
Gross-Pitaevskii equation and used it to investigate the dynamics and
expansion of the condensate in a one-dimensional optical lattice.
Finally, Kr\"{a}mer {\it et al.}~\cite{Kraemer2002a}
used hydrodynamic equations to study the low-lying 
collective modes of a harmonically trapped
Bose-Einstein condensate in the presence of a one-
or two-dimensional optical lattice. 
In particular, they showed that the effect of the lattice is to renormalize
the interaction coupling constant and introduce an effective
mass that accounts for the different inertia along the lattice potential.
With these changes it is for instance 
possible to apply the
results for harmonically trapped condensates obtained by
Stringari~\cite{Stringari1996a}.

The study by Kr\"{a}mer {\it et al.} is somewhat related to ours. 
The most important difference, however, is that 
Kr\"{a}mer {\it et al.} are interested in
different modes than we are. They deal with low-energy excitations
along the long axis of the cigar-shaped condensate, i.e.,   the longitudinal
modes. In the absence of a magnetic
trap they correspond to the Bogoliubov modes with the familiar
phonon spectra at large wave lengths. In the presence of a harmonic trap
the spectrum becomes discrete and the lowest energetic modes are 
the center-of-mass mode and the (longitudinal) quadrupole-mode. The transverse 
modes we are interested in
have superfluid flows orthogonal to the long axis of the condensate.
Moreover, in the $z$-direction along the lasers of the optical lattice
the condensate is for simplicity treated as completely periodic. 
As we see later, 
we take in first instance 
the atom number in each two-dimensional (quasi)condensate
to be constant and equal in every site. This
removes the Bogoliubov modes from the excitation spectrum, 
since they correspond to a density modulation 
propagating along the $z$-axis. If desired, they can however be easily
incorporated into our approach.
Indeed, at the end of the paper we briefly discuss the corresponding sound mode
when the atom number in each two-dimensional (quasi)condensate is allowed
to fluctuate.

There are several experiments on condensates in a one-dimensional 
optical lattices
~\cite{Greiner2001a,Cataliotti2001a,Burger2001a,Cristiani2002a,
Morsch2002a,Fort2003a}, 
but none of these address the problem we consider in this paper.
Fort {\it et al.}~\cite{Fort2003a} measured longitudinal excitation frequencies
of the condensate in the presence of a one-dimensional 
optical lattice and this is the
experimental paper most closely related to our work.
In particular, for the breathing mode 
Fort {\it et al.}~\cite{Fort2003a} 
do not report any dependence on the lattice depth
and this is in agreement with our result as well as with the result
of Kr\"{a}mer {\it et al.}~\cite{Kraemer2002a}.
While we do not
predict dependence on the lattice depth, we do predict 
dependence on the modulation of the excitation along the $z$-axis.
In particular, we expect changes in the eigenfrequency if the sites
are out of phase with each other and these changes can be large
for typical experimental parameters. So far such phenomena have not
been probed experimentally.

The paper is organized as follows. In Sec.~\ref{sec:model} we derive
the theory used in this paper. In Sec.~\ref{sec:novortex} we apply
this theory to a Bose-Einstein condensate without vortices and
calculate the dispersion relation of the monopole and quadrupole-modes
in the presence of a lattice. In Sec.~\ref{sec:vortex} we proceed 
by repeating similar calculation for 
the  vortex state of the Bose-Einstein condensate. We end with a discussion
of our results in Sec.~\ref{sec:conclusions}.

\section{Gross-Pitaevskii theory}
\label{sec:model}
Our starting point is a cigar-shaped Bose-Einstein condensate trapped by
the potential
\beq
\label{magnetic_trap}
V({\bf r})=\frac{M}{2}\left(\omega_r^2r^2+\omega_z^2z^2
\right),
\enq
where $\omega_r$ and $\omega_z$ are the radial and axial 
trapping frequencies, respectively, and $M$ is
the atomic mass. As we assume a 
cigar-shaped trap, we further have that $\omega_z\ll \omega_r$. 
The condensate also experiences an one-dimensional optical lattice
\beq
\label{lattice_potential}
V_0({\bf r})=V_0\sin^2\left(\frac{2\pi z}{\lambda}\right),
\enq
where $V_0$ is the lattice depth and $\lambda$ is the wave length of the
laser-light. We assume that the lattice is deep enough so that it dominates
over the magnetic trapping potential in the $z$-direction. 
When this is true and the number of lattice sites is large, 
i.e., $\lambda\ll l_z=\sqrt{\hbar/M\omega_z}$,
we can in first instance
ignore the magnetic trapping potential in the $z$-direction.

The lattice potential splits the condensate into $N_s$
two-dimensional (quasi)condensates with a pancake shape.
We assume that the lattice is sufficiently deep
such that its depth is larger than the chemical potential
of the two-dimensional (quasi)condensate~\cite{comment2003a}. 
Using a Thomas-Fermi approximation
for the two-dimensional (quasi)condensates we obtain a lower boundary
that can be expressed as
\beq
V_0\gg 2^{9/7}\left[N a 
\left(\frac{M\omega_r}{\hbar\lambda^2}\right)^{1/4}\right]^{4/7}
\hbar\omega_r,
\enq
where $N$ is the number of atoms per site and
$a$ is the three-dimensional scattering length.
As a numerical example, we take an $^{87}{\rm Rb}$ condensate in a trap
with a radial trapping frequency 
$\omega_r/2\pi=100 {\rm Hz}$ and a laser-light 
wave length of $\lambda=800 {\rm nm}$.
When the atom number in each site is between $100$ and $1000$, the
lower bound on the trap depth $V_0$ is between 
$0.05 E_r$ and $0.19 E_r$, where 
$E_r=\hbar^2\left(2\pi/\lambda\right)^2/2M$ is the 
recoil energy of an atom that absorbed one photon from the laser beam.

Although we are interested in a deep lattice, we consider here
only the case that there is still full coherence 
across the condensate array. Specifically this means that
the lattice potential should not be so deep as to induce a Mott-insulator
transition. Typically the required lattice depth to reach
the Mott-insulator transition in a three-dimensional lattice with 
a filling factor of one is
of the order of $10 E_r$. In a one-dimensional lattice
the number of atoms in each lattice site is typically much larger 
than in a three-dimensional lattice and the transition into 
the insulating state requires a much deeper lattice. 
In mean-field theory the Mott-insulator transition in such a system  occurs
when $U_R>8NJ$~\cite{vanOosten2003a}, 
where $U_R$ and $J$ are respectively the characteristic renormalized
interaction and
hopping parameters of the effective single-mode 
Bose-Hubbard model with Hamiltonian
\beq
\hat{H}=-J\sum_{<i,j>}\hat{b}_j^\dagger\hat{b}_i %-\mu\sum_i \hat{n}_i
+\frac{U_R}{2}\sum_i \hat{n}_i\left(\hat{n}_i-1\right).
\enq
Using the same numerical values as in the previous paragraph,
we estimate the critical lattice depth for the Mott-insulator transition
to be between $56 E_r$ and  $82 E_r$, when the number of atoms
in each site is again between $100$ and $1000$.
To the best of our knowledge 
the Mott-insulator transition in a one-dimensional optical lattice
has not yet been observed. 

We use trap units from now on, i.e., the unit of energy is $\hbar\omega_r$,
 the unit of time is $1/\omega_r$,
and the unit of length is $l_r=\sqrt{\hbar/M\omega_r}$.
The Gross-Pitaevskii energy functional, which describes the system
at low temperatures, is then
\begin{eqnarray}
\label{fullH}
E\left[\Psi^*,\Psi\right]=
\int d{\bf r} \left\{-\frac{1}{2}\Psi^*({\bf r})\nabla^2\Psi({\bf r})+
\left[\frac{1}{2}\left(x^2+y^2\right)+\frac{V_0({\bf r})}{\hbar\omega_r}
+\frac{T^{2B}}{2}
|\Psi({\bf r})|^2\right]|\Psi({\bf r})|^2\right\},
\end{eqnarray}
where $T^{2B}$ is the two-body $T$-matrix. In the above units the latter is
related to the three-dimensional 
$s$-wave scattering length $a$ through $T^{2B}=4\pi a/l_r$.

For a deep lattice potential it is natural to expand 
the condensate wave function in terms of wave functions that
are well localized in the sites. More precisely, we expand
\beq
\label{localized_ansatz}
\Psi\left({\bf r}\right)=\sum_n w\left(z-z_{n}\right)
\Phi_n\left(x,y\right),
\enq
where $n$ labels the lattice sites and $z_n=\lambda n/2 l_r$ 
is the position of the
$n$th site. For now we do not specify the 
wave functions $\Phi_n\left(x,y\right)$ of the
two-dimensional (quasi)condensates, but for the wave function
in the $z$-direction, $w\left(z\right)$, we use the ground-state
wave function of the harmonic approximation to the lattice potential near the
lattice minimum. This harmonic trap has the frequency 
\beq
\omega_L=\frac{2\pi}{\lambda}\sqrt{2V_0/M}
\enq
and the wave function $w(z)$ is thus given by
\beq
\label{z_profile}
	w(z)=\frac{1}{\pi^{1/4}\sqrt{l_L}}\; %\cdot \frac{1}{\sqrt{l_L}}
\exp\left(-\frac{z^2}{2 l_L^2}\right),
\enq
where $l_L=\sqrt{\hbar/M\omega_L}$.

Substituting the above ansatz into the energy functional
and ignoring all but the nearest neighbor interactions, we get the 
energy functional
\begin{eqnarray}
\label{2D_Hamiltonian}
E\left[\Phi^*,\Phi\right]
&=&\sum_n \int d^2r\left\{-\frac{1}{2}\Phi_n^*(x,y)\nabla^2\Phi_n(x,y)
+\left[\frac{1}{2}\left(x^2+y^2\right)+\frac{U_{2D}}{2}|\Phi_n(x,y)|^2
\right]|\Phi_n(x,y)|^2\right.\nonumber\\
&-&\left.J\sum_{<n,m>}\int d^2r \Phi^*_m(x,y)\Phi_n(x,y)\right\},
\end{eqnarray}
where $\langle n,m \rangle$ indicates nearest neighbors, and
\beq
U_{2D}=T^{2B}\int dz |w(z)|^4=4\sqrt{\frac{\pi}{2}}
\left(\frac{a}{l_L}\right)
\enq
is the two-dimensional coupling strength. Moreover, $J$ 
is the strength of the Josephson coupling between neighboring sites
and we have
\begin{eqnarray}
\label{Jdefinition}
J=-\int dz w^*(z)\left[
-\frac{1}{2}\frac{\partial^2}{\partial z^2}+\frac{V_0(z)}{\hbar\omega_r}
\right]
w(z+\lambda/2\,l_r).
\end{eqnarray}
With these assumptions $J$ is a time-independent 
experimentally defined parameter. 
Approximating the lattice potential near its maximum by an upside-down 
parabolic potential we can
calculate the Gaussian integral, with the result
\beq
J=\frac{1}{8\pi^2}\left(\frac{\omega_L}{\omega_r}\right)^2
\left(\frac{\lambda}{l_r}\right)^2\left[\frac{\pi^2}{4}-1\right]
e^{-\left(\lambda/4\,l_L\right)^2}.
\enq

The energy functional in Eq.~(\ref{2D_Hamiltonian}) 
is now almost two-dimensional. The third dimension
is visible only in the last term that describes the coupling between
neighboring layers. The energy is characterised by two 
parameters $U_{2D}$ and $J$, both of which are experimentally tunable.
The importance of the on-site interaction 
term proportional to $U_{2D}$ can be enhanced by increasing 
the number of particles in the sites or by making the lattice deeper.
Deepening the lattice also decreases the strength of the Josephson coupling 
$J$ and makes the sites
more independent. It should be noticed that while $J$ is tunable, it is
always positive. Physically this means that there is always
an energetic penalty for having a phase difference between sites.

%\asjhdsa
\section{Excitations of the condensate ground state}
\label{sec:novortex}
In this section we study the transverse excitations of the
ground-state of the stack of two-dimensional (quasi)condensates.
Using a Gaussian ansatz for the wave functions of the 
(quasi)condensates, we solve the dispersion relations
for the monopole and the quadrupole-modes analytically.
In Sec.~\ref{sec:novortex_nolattice} we introduce the
Gaussian ansatz and solve the excitations for an individial
two-dimensional (quasi)condensate. In Sec.~\ref{sec:novortexlattice} we proceed
to calculate the band structure of the monopole and
quadrupole-modes in the optical lattice. These sections also include
technical details about the calculations. Such details are
not repeated in Sec.~\ref{sec:vortex} when we consider the vortex state. 

\subsection{Excitations for a single two-dimensional (quasi)condensate}
\label{sec:novortex_nolattice}
To account for the monopole and quadrupole modes of the
two-dimensional (quasi)condensates in every site, we use
a general Gaussian ansatz for the wave functions, i.e.,
\beq
\label{gaussianansatz}
\Phi_n(x,y,t)=C_n(t)
\exp\left[-\frac{1}{2}\left(B_{xx,n}(t)x^2+B_{yy,n}(t)y^2+2B_{xy,n}(t)xy\right)
\right].
\enq
All three variational parameters 
$B_{ij,n}(t)\equiv B_{ij,n}'(t)+iB_{ij,n}''(t)$ 
are complex. From now on we always use a prime 
to denote the real part of a complex quantity and a 
double prime to denote its imaginary part.
The wave functions are normalized to the number of particles $N$
at the site and therefore
\beq
C_n(t)=\sqrt{\frac{N}{\pi}}\left(B_{xx,n}'(t)B_{yy,n}'(t)
-B_{xy,n}'(t)^2\right)^{1/4}.
\enq
As we fix the number of particles in every site, we are excluding
the Bogoliubov modes propagating along the $z$-axis. 
It is, however, not difficult to account also for these modes as we
show later on.
The equations of motion for the variational parameters can be derived from 
the Lagrangian
\beq
L\left[\Phi^*,\Phi\right]=\frac{i}{2}\int d^2r
\left(\sum_n \Phi_n^*(x,y,t)\frac{\partial \Phi_n(x,y,t)}{\partial t}-
\Phi_n(x,y,t)\frac{\partial \Phi_n^*(x,y,t)}{\partial t}
\right)-E\left[\Phi^*,\Phi\right].
\enq

Let us first investigate the behaviour of an individual two-dimensional
(quasi)condensate.
Without the interlayer coupling the part of the 
Lagrangian quadratic in the deviations
$\epsilon_{ij}(t)$ turns out to be equal to
\begin{eqnarray}
\label{in_plane_Lagrangian_no_vortex}
\frac{L}{N}&=&-\frac{1}{4B_0^2}\left(\epsilon_{xx}'\dot{\epsilon}_{xx}''+
\epsilon_{yy}'\dot{\epsilon}_{yy}''+2\epsilon_{xy}'\dot{\epsilon}_{xy}''
\right)+\frac{U}{B_0}\left[
\frac{\left(\epsilon_{xx}'-\epsilon_{yy}'\right)^2}{8}+
\frac{\epsilon_{xy}'^2}{2}\right]+
\nonumber\\
&-&\left(\frac{1}{2B_0^3}+\frac{1}{2B_0}\right)\left(
\epsilon_{xx}'^2+\epsilon_{yy}'^2+\epsilon_{xx}'\epsilon_{yy}'
+\epsilon_{xy}'^2\right)+\left(\frac{1}{4B_0^3}+\frac{3}{4B_0}\right)
\left(\epsilon_{xx}'+\epsilon_{yy}'\right)^2\nonumber\\
&-&\frac{1}{4B_0}\left[\left(\epsilon_{xx}'+\epsilon_{yy}'\right)^2
+2\epsilon_{xx}'\epsilon_{yy}'+\epsilon_{xx}''^2
+\epsilon_{yy}''^2+2\epsilon_{xy}''^2
\right],
\end{eqnarray}
where 
\beq
U=\frac{N}{\sqrt{2\pi}}\left(\frac{a}{l_r}\right)
\sqrt{\frac{\omega_L}{\omega_r}}.
\enq
We also defined the equilibrium solution of $B_{ii}(t)$
as $B_0$. Hence, 
$B_{ij}(t)=B_0\delta_{ij}+\epsilon_{ij}(t)$. 
We also suppressed the site index $n$. 
The equilibrium solution $B_0$ is given by minimizing the zeroth-order
term of the energy, i.e.,
\beq
E_0=\frac{1}{2B_0}+\frac{B_0}{2}+UB_0
\enq
with the result
\beq
\label{B0_novortex}
B_0=\sqrt{\frac{1}{1+2U}}.
\enq
In Eq.~(\ref{in_plane_Lagrangian_no_vortex}) we show only
the part relevant for the dynamics and we ignored the zeroth-order term
whose minimisation lead to the result in Eq.~(\ref{B0_novortex}).

We are now in a position to find the frequencies for the collective excitations
we are interested in.
Let us start with the monopole $m=0$ 
mode, which is alternatively also called the breathing mode. 
For the monopole-mode we can set 
$\epsilon_{xx}=\epsilon_{yy}=\epsilon$ and $\epsilon_{xy}=0$.
With this choice the Lagrangian is greatly simplified to
\beq
\label{Lagrangian_novortex_breathing}
\frac{L}{N}=-\frac{1}{2B_0^2}\left[\epsilon'\dot{\epsilon}''-
\frac{1}{B_0}\epsilon'^2-B_0\epsilon''^2\right].
\enq
The equations of motion for $\epsilon'$ and $\epsilon''$ 
are the Euler-Lagrange equations that result in
%\beq
%\frac{d}{dt}\frac{\partial L}{\partial \dot{\epsilon}^\alpha}-
%\frac{\partial L}{\partial \epsilon^\alpha}=0
%\enq
two first-order differential equations
\begin{eqnarray}
\dot{\epsilon}'+2B_0\epsilon''=0\nonumber\\
-\dot{\epsilon}''+\frac{2}{B_0}\epsilon'=0.
\end{eqnarray}
These equations can be cast 
into a single second-order differential equation for $\epsilon'$
\beq
\ddot{\epsilon}'=-4\epsilon',
\enq
which describes sinusoidal oscillation with a frequency $2$.
The frequency of the monopole-mode is therefore
\beq
\label{breathing}
\omega_0=2
\enq
and it is independent of the strength of interactions.
This is in agreement with previous results~\cite{Kagan1997a}.

The quadrupole $m=\pm 2$ modes are captured by the choice
$\epsilon_{xx}=-\epsilon_{yy}=\epsilon$. 
We then have just two (complex) variational parameters,
$\epsilon$ and $\epsilon_{xy}$.
In the Lagrangian in Eq.~(\ref{in_plane_Lagrangian_no_vortex})
there are no terms that couple $\epsilon$ to $\epsilon_{xy}$.
Therefore, the dynamics of these parameters
separates and both turn out to have the same oscillation frequency.
Above we gave the necessary technical details in the derivation of the
monopole-mode frequency. As the quadrupole-mode frequency can be dealt
with in a similar fashion, we simply give the result.
The quadrupole-mode frequencies are given by
\beq
\label{quadrupole}
\omega_{\pm 2}=\sqrt{2+2B_0^2}.
\enq
For the ideal-gas $B_0=1$
and the quadrupole frequency is again $2$. In the Thomas-Fermi limit
$B_0$ tends to zero and the quadrupole frequencies approach 
$\omega_{\pm 2}\rightarrow\sqrt{2}$. Again this result is as 
expected~\cite{Jin1996a,Stringari1996a,Edwards1996a}.
Our treatment also captures the scissors mode~\cite{Khawaja2001c}, but in 
the axial symmetric case we are considering here the scissors mode
turns out to be degenerate with the quadrupole-mode.

Incidentally, it should be remembered that the degeneracy of the quadrupole
modes 
is lifted in a rotating trap. If the trap is rotating with frequency
$\Omega$ around the $z$ axis 
we should include a term $-\Omega \langle L_z\rangle$ into
the energy functional, where $\langle L_z\rangle$ is the expectation
value of the angular momentum component in the $z$-direction.
The angular momentum of the equilibrium solution is zero and the
new term will only contribute in second order. The new contribution
to the energy is
\beq
-\Omega \langle L_z\rangle=\frac{\Omega}{B_0^2}\left[
\epsilon_{xy}'\epsilon''-\epsilon_{xy}''\epsilon'\right].
\enq
This term couples the dynamics of $\epsilon$ and $\epsilon_{xy}$, but
the resulting $2\times 2$ matrix problem is easy to solve.
The quadrupole-mode frequencies in a rotating trap are 
\beq
\omega_{\pm 2}=\sqrt{2}
\left[\left(1+B_0^2\right)^{1/2}\pm\sqrt{2}\Omega\right].
\enq
From this result it is clear that the quadrupole
mode with $m=-2$ becomes thermodynamically unstable when 
$\Omega>\sqrt{\left(1+B_0^2\right)/2}$.
This result corresponds to the Landau-criterion for the quadrupole-modes,
and 
has been shown to play an important role in the nucleation of
vortices~\cite{Madison2001a,Hodby2002a,Feder1999a,Recati2001a,Sinha2001a}.

\subsection{Influence of the lattice on the excitation frequencies}
\label{sec:novortexlattice}
We are now in the position to discuss the influence of the lattice
potential. To make progress we must determine the coupling integral
\beq
I_{mn}=\int d^2r \Phi^*_m(x,y)\Phi_n(x,y)
\enq
to a sufficient accuracy. This will contribute to the 
energy a Josephson coupling
\beq
H_{J}=-J\sum_{\langle n,m\rangle} I_{mn}',
\enq
where the $\langle n,m\rangle$ indicates nearest neighbors. 
Here the imaginary part of $I_{mn}$ is not relevant since
its contribution to the energy 
vanishes when the sum over the nearest-neighbors is calculated. 
For the monopole-mode we get up to second order in
the deviations  the result
\beq
\label{monopoleI}
\frac{I_{mn}'}{N}= %\begin{array}{l}
1-\frac{1}{8B_0^2}\left(\epsilon_{n}'^2+\epsilon_{m}'^2+
2\epsilon_{n}''^2+2\epsilon_{m}''^2\right)
+\frac{1}{4B_0^2}\epsilon_{n}'\epsilon_{m}'
+\frac{1}{2B_0^2}\epsilon_{n}''\epsilon_{m}''  %\right.
%\end{array}
\enq
and for the quadrupole-mode we have
\begin{eqnarray}
\label{quadrupoleI}
\frac{I_{mn}'}{N}&=&%\left\{\begin{array}{l}
1-\frac{1}{8B_0^2}\left(
|\epsilon_n|^2+|\epsilon_m|^2+|\epsilon_{xy,n}|^2+|\epsilon_{xy,m}|^2\right)+
\nonumber \\
&+&\frac{1}{4B_0^2}\left(\epsilon_{n}'\epsilon_{m}'+
\epsilon_{n}''\epsilon_{m}''+\epsilon_{xy,n}'\epsilon_{xy,m}'+
\epsilon_{xy,n}''\epsilon_{xy,m}''\right).
%\quad\mathrm{quadrupole}
%\end{array}
%\right.
\end{eqnarray}
In these formulae the first subindex of $\epsilon_{xy,n}$
identifies the variational parameter in question and 
the second indicates the lattice site. For identical nearest-neighbor
wave functions the overlap integral $I_{mn}$ should be exactly $N$,  
which is indeed the case in both equations (\ref{monopoleI}) and
(\ref{quadrupoleI}).

Some terms in Eqs.~(\ref{monopoleI}) and (\ref{quadrupoleI}) 
are purely on-site, 
but terms of the type $\epsilon_{n}\epsilon_{m}$
are not. This complication is remedied by going to Fourier-space.
We define the Fourier transform in such a way that the function $f_n$ in 
coordinate space is expressed in terms of its transform $f_k$ as
\beq
\label{ftransform}
f_{n}=\frac{1}{\sqrt{N_s}}\sum_{k=-\frac{2\pi}{\lambda}\left(1-\frac{1}{N_s}
\right)}^{\frac{2\pi}{\lambda}\left(1-\frac{1}{N_s}
\right)} \exp\left[
ikz_n\right] f_{k}.
\enq
Here $N_s$ is the number of lattice sites which we, for notational
convenience, assume to be an odd number. Moreover, 
$k$ is the wave number and the lattice spacing is $d=\lambda/2$.

First we transform the diagonal terms in the Lagrangian. For example
\beq
\sum_n f_{n}^2=\sum_n\frac{1}{N_s}\sum_{k,k'} f_kf_{k'}\exp\left[
iz_n\left(k+k'\right)\right]=\sum_k f_kf_{-k}=\sum_k |f_k|^2,
\enq
where the sum over the lattices sites $n$ gave the Kronecker delta 
$\delta_{k',-k}$ which removed one of the momentum sums. The
last step is a result of the fact that $f_n$ was a real function, so
$f_k^*=f_{-k}$.
Nearest-neighbor terms are somewhat more complicated. As an example
\beq
\sum_{<n,m>} f_nf_m=\frac{1}{2N_s}\sum_{n}\sum_{k,k'}f_kf_{k'}\left[
\exp\left(i\left(kz_n+k'z_{n+1}\right)\right)+
\exp\left(i\left(kz_n+k'z_{n-1}\right)\right)\right].
\enq
We can perform the sum over $n$ and get
\beq
\sum_{<n,m>} f_nf_m=\sum_{k,k'} \cos\left(k'\lambda/2\right)
f_kf_{k'}\delta_{k',-k}=\sum_k \cos\left(k\lambda/2\right) |f_k|^2.
\enq
In Fourier-space the Josephson coupling 
$H_{J}$ generally thus introduces factors of
$\cos\left(k\lambda/2\right)-1$
into the Lagrangian. 

Now that we know how to transform to Fourier-space,
we can proceed to derive equations of motion for each
value of the wave vector $k$. Since two different values
of the wave vectors do not couple,
this is not technically any more 
complicated than our previous treatment of an individual (quasi)condensate.
The equations for each wave vector can be solved separately.
We demonstrate this again for the simplest case, namely the
breathing mode. Let the Fourier transform of $\epsilon_{n}$ be
$\epsilon_{k}$.
In  Fourier space the Lagrangian for the
breathing mode is 
\begin{eqnarray}
L&=&-\frac{1}{2B_0^2}\left\{\sum_k \epsilon_{k}'^*\,\dot{\epsilon}_{k}''+
\left[\frac{1}{B_0}-J\left(\cos\left(\frac{k\lambda}{2}\right)-1\right)
\right]|\epsilon_{k}'|^2+\right.\nonumber\\
&+&\left.\left[B_0-2J\left(\cos\left(\frac{k\lambda}{2}\right)-1\right)\right]
|\epsilon_{k}''|^2\right\}.
\end{eqnarray}
Keeping in mind that $\epsilon_{-k}=\epsilon_{k}^*$
we get equations of motion for $\epsilon_{k}'$ and $\epsilon_{k}''$.
For example by considering the variation of the Lagrangian
with respect to
$\epsilon_{-k}''$  we get
%\beq
%\frac{d}{dt}\frac{\partial L}{\partial \dot{\epsilon}_{-k}''}-
%\frac{\partial L}{\partial \epsilon_{-k}''}=0
%\enq
%gives us
\beq
\label{Lag_eq1}
\dot{\epsilon}_{k}'-2
\left[B_0-2J\left(\cos\left(\frac{k\lambda}{2}\right)-1\right)\right]
\epsilon_{k}''=0
\enq
and by considering variations with respect to $\epsilon_{-k}'$ 
we get the differential equation for $\epsilon_{k}''$
\beq
\label{Lag_eq2}
\dot{\epsilon}_{k}''+2\left[\frac{1}{B_0}-
J\left(\cos\left(\frac{k\lambda}{2}\right)-1\right)\right]
\epsilon_{k}'=0.
\enq
The dispersion relation for the monopole-mode can now be simply
read out from this pair of equations. The quadrupole-modes
can be dealt with in the same way although the equations are
somewhat longer.

\begin{figure}
\includegraphics[width=\columnwidth]{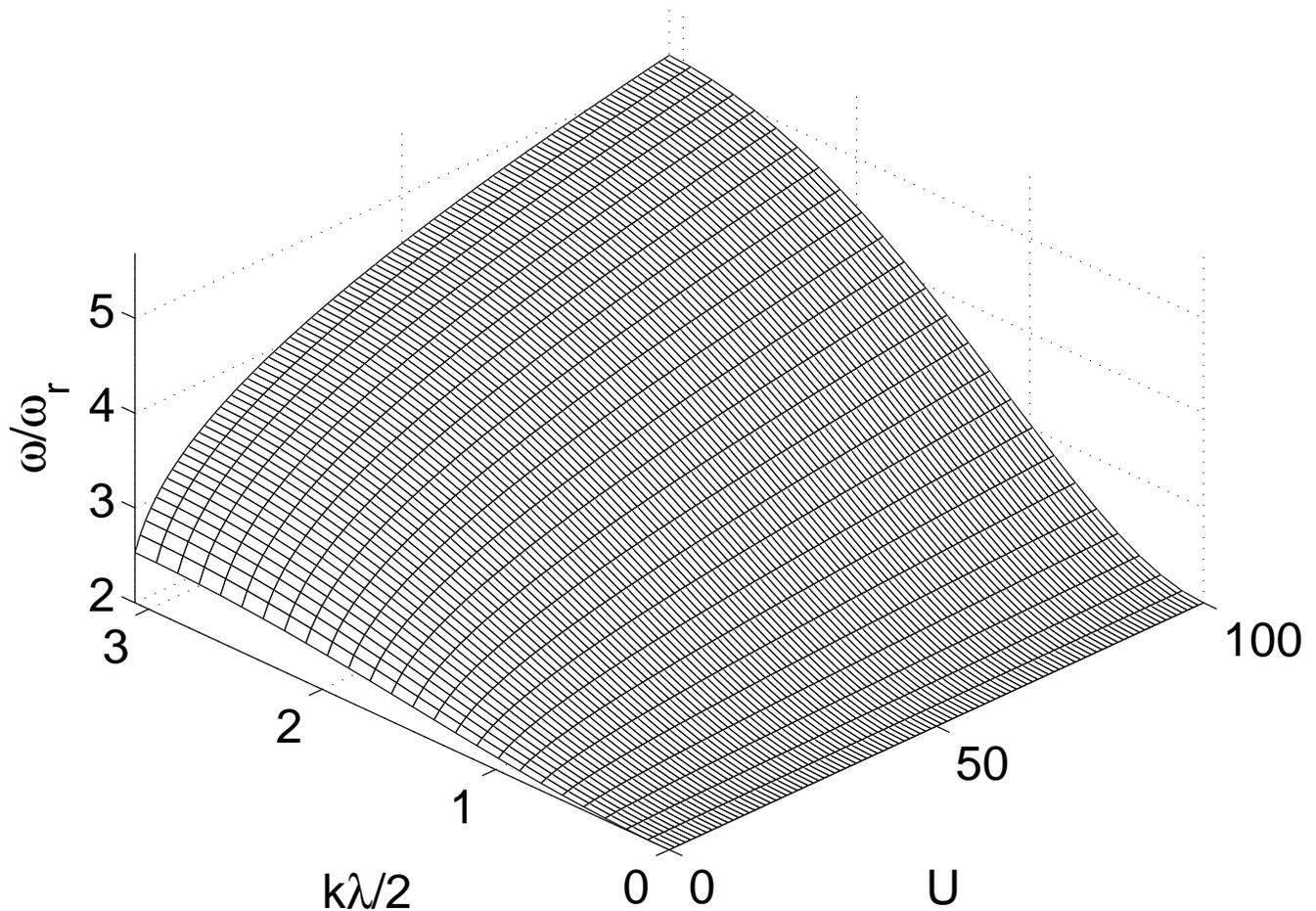}
%\includegraphics[width=10.0cm]{splitting.eps}
%\vspace*{1cm}
\caption[]{monopole-mode frequency as a function of $k$ and $U$, when
$J=0.05$. The surface in this figure was calculated using 
Eq.~(\ref{monopolemode}).
\label{fig:monopolemode}}
\end{figure}

For convenience we assume that the contribution 
from terms proportional to $J^2$ are very small. With this
simplification we get  
the dispersion relations for the monopole and quadrupole-modes 
\beq
\label{monopolemode}
\omega_0(k)=2\left[1-J\left(B_0+\frac{2}{B_0}\right)
\left(\cos\left(\frac{k\lambda}{2}\right)-1\right)
\right]^{1/2}
\enq
\beq
\label{quadrupolemode}
\omega_{\pm 2}(k)=\sqrt{2}\left[1+B_0^2-J\left(3B_0+\frac{1}{B_0}\right)
\left(\cos\left(\frac{k\lambda}{2}\right)-1\right)
\right]^{1/2}.
\enq
We emphasize that our results where terms under the
square root proportional 
to $J^2$ are ignored should be used with some caution.
The terms proportional to $J^2$ are not always negligible compared to the
other contributions. In particular
if the trap depth or the on-site number of particles
is small, there is a range of experimentally relevant parameter values
where terms proportional to $J^2$ can be relatively large and should be
included. They will not change the qualitative behavior of the dispersion
relations, but can affect quantitative results. While we choose
to work in the regime where terms proportional to $J^2$ are small,
it is not difficult to include these missing terms. For example,
Eqs.~(\ref{Lag_eq1}) and ~(\ref{Lag_eq2}) show that the exact frequency
for the monopole-mode obeys
\beq
\omega_{0}^2(k)=4
\left[B_0-2J\left(\cos\left(\frac{k\lambda}{2}\right)-1\right)\right]
\left[\frac{1}{B_0}-J\left(\cos\left(\frac{k\lambda}{2}\right)-1\right)\right].
\enq
In Fig.~\ref{fig:monopolemode} we show the dispersion relation for
the monopole-mode as a function of $k$ and $U$.

In the limit of long wavelengths it is permissible to expand
the cosine factors. The excitation has then the same dispersion 
as that of a
free-particle $\Delta+\hbar^2k^2/2m^*$ with some effective mass $m^*$
and a gap $\Delta$.
For the monopole-mode we, therefore, predict an effective mass
\beq
m_0^*=\frac{4B_0}{J\left(B_0^2+2\right)}
\,\left(\frac{\hbar}{\omega_r\lambda^2}\right)
\enq
and for the quadrupole-mode we get
\beq
m_{\pm 2}^*=
\frac{4B_0\sqrt{2}}{J\left(3B_0^2+1\right)}
\,\left(\frac{\hbar}{\omega_r\lambda^2}\right).
\enq
It is quite interesting to observe that the 
effective masses of different modes are different. 
In particular, the effective mass 
of the quadrupole-mode is bigger than the 
effective mass of the monopole-mode. This can be understood by
considering the overlap integral between neighboring sites.
For the monopole-mode the coupling between the nearest neighbors 
is to a large extent determined by the integral 
\beq
\int d^2r |\Phi_0(x,y)|^2 (x^2+y^2)^2,\nonumber
\enq
where $\Phi_0(x,y)$ is the equilibrium wave function. 
In contrast, for the quadrupole-mode the coupling is determined by
the integral
\beq
\int d^2r |\Phi_0(x,y)|^2 (x^2-y^2)^2\nonumber.
\enq
It is clear that the latter integral is smaller than the first
one. As the effective mass is inversily proportional to the
strength of the nearest-neightbour coupling, the quadrupole-mode
therefore has a bigger effective mass. 

The fact that the dispersion relation is similar to the free-particle
dispersion relation is reflected in the dynamics. 
A sinusoidal modulation 
moves with velocity $v\simeq\hbar\langle k\rangle/m^*$ in the $z$-direction. 
In a finite system a
pure sinusoidal modulation is not possible and the excitation corresponds 
to a wavepacket centered around $\langle k\rangle$ and
with some nonzero width $\Delta k$. If the system
is large enough, i.e., much bigger than  
$2\pi/\Delta k$, the width of the
packet can be small and one should be able to observe such propagation
before the excitation hits the outer edge of the condensate.

More generally we can expand the dispersion relation around any
value of the wave vector. 
In terms of a function $C(J,B_0)$ that
depends on the mode in question, the excitation energy up to lowest-order
in $J$ looks like
\beq
\omega(k)=\omega(0)+
C\left(J,B_0\right)\left(\cos\left(\frac{k\lambda}{2}\right)-1\right).
\enq
Expanding this expression around $k_0$ we get
\beq
\omega(k)\approx \omega(k_0)-C\left(J,B_0\right)\left[\frac{\lambda}{2}
\sin\left(\frac{k_0\lambda}{2}\right)\left(k-k_0\right)
+\frac{\lambda^2}{8}\cos\left(\frac{k_0\lambda}{2}\right)
\left(k-k_0\right)^2
\right].
\enq
When $k_0=0$ we get the results for the effective masses we presented
earlier, but some special cases are also of interest. In particular, when
$k\lambda/2=\pi$ we obtain 
the same expansion as with $k=0$, but the constant in front of 
$\left(k-k_0\right)^2$ has a negative sign. In this regime the effective
mass is therefore negative. In the regime of a negative effective
mass one encounters modulational instabilities 
as discussed in Refs.~\cite{Wu2001a,Konotop2002a,Baizakov2002a}.

\section{Excitations of the vortex state}
\label{sec:vortex}
In this section we consider a 
system of weakly-coupled two-dimensional 
(quasi)condensates that has a vortex  piercing through
the center of 
each (quasi)condensate. For such a system our 
earlier ansatz in Eq.~(\ref{gaussianansatz}) is inadequate. 
For an individual condensate it is known that the presence of a vortex
should not change the dispersion of the monopole-mode, but it
will lift the degeneracy of the quadrupole-modes. Physically 
this is due to the fact that the quadrupole excitation, depending
on the sign of the quantum number $m$, travels either in the
same direction of the superfluid flow or opposite to it.
As the monopole-mode is easier to tackle than the quadrupole-modes,
we start with that in Sec.~\ref{sec:monopole_vortex}. In 
Sec.~\ref{sec:quadrupole_vortex} we solve for the quadrupole-modes
of an individual two-dimensional (quasi)condensate and
in Sec.~\ref{sec:vortexquadrupolelattice} we include also
for the quadrupole-mode the optical
lattice into our discussion.

\subsection{monopole-mode in the presence of a vortex}
\label{sec:monopole_vortex}
The vortex state
has a superfluid flow around the vortex core. This flow diverges in the
core and for this reason the density of the condensate must vanish
in the vortex core. The simplest ansatz having these two desired properties
is (in polar coordinates)
\beq
\label{vortex_monopole_ansatz}
\Phi_n(r,\phi)\propto r\exp\left[i\phi\right]
\exp\left[-\frac{B_0r^2}{2}\right]
\exp\left[-\frac{\epsilon_{n}(t)r^2}{2}\right].
\enq
The ansatz is almost the same as in the previous section for the monopole-mode
of the state
without a vortex. The only differences are the first two factors that give
the vortex the properties we were after. The size of the vortex core region 
in Eq.~(\ref{vortex_monopole_ansatz}) is about $1/\sqrt{B_0}$ and it does not 
diminish as the number of particles is
increased. This is in principle incorrect, since 
the length scale for the
vortex core size is set by the coherence length and 
the coherence length in the center of the condensate gets smaller as the
number of particles is increased.
We expect that this
unphysical behaviour close to the vortex core is not relevant to the
physics of the collective modes at hand. In the end of the calculations
we can reproduce the known results
for the individual pancake to a good accuracy and thus
our expectations are indeed well justified.

Using similar techniques as for the condensate without
a vortex, 
we calculate the monopole-mode of a condensate with a vortex as
\beq
\label{monopolemode_vortex}
\omega_0(k)=2\left[1-3J\left(B_0+\frac{1}{B_0}\right)
\left(\cos\left(\frac{k\lambda}{2}\right)-1\right)
\right]^{1/2},
\enq
where we have again assumed that $J^2$ terms under the square root 
can be ignored.
The equilibrium solution is now given by
\beq
\label{equilibrium_vortex}
B_0=\sqrt{\frac{2}{2+U}}.
\enq
The result is similar to Eq.~(\ref{monopolemode}), but the constant
in front of the cosine term is different indicating a difference in the
effective mass. For the monopole-mode in the presence of a vortex we get
\beq
m_{0,v}^*=\frac{4B_0}{3J\left(B_0^2+1\right)}
\,\left(\frac{\hbar}{\omega_r\lambda^2}\right).
\enq
We can see that the effective mass of the monopole-mode of the vortex state
is somewhat smaller than the effective mass in the absence 
of a vortex. This can be understood by comparing the relevant 
overlap integrals
for the wave functions with and without the vortex.
Since the (quasi)condensate wave function with a vortex is more extended
than without a vortex, the strength of the
nearest-neighbor coupling is increased and, therefore, the effective mass
is reduced.

\subsection{quadrupole-modes of the single pancake 
in the presence of a vortex}
\label{sec:quadrupole_vortex}
As we mentioned before the quadrupole-modes are more complicated. For the
quadrupole-modes we use the ansatz
\begin{eqnarray}
\label{quadrupole_ansatz}
\Phi_n(r,\phi)&\propto& 
r \exp\left[i\phi\right]\exp\left[-\frac{B_0r^2}{2}\right]
\exp\left[-\epsilon\frac{\left(x^2-y^2\right)}{2}-\epsilon_{xy}xy\right]
\left[1+\alpha \exp\left[-2i\phi\right]\right]\nonumber\\
&\simeq&r\exp\left[i\phi\right]\exp\left[-\frac{B_0r^2}{2}\right]
\left[1+
\cos\left(2\phi\right)\left(\alpha-\frac{\epsilon}{2}r^2\right)
\right.\nonumber\\
&-&\left.\sin\left(2\phi\right)\left(i\alpha+\frac{\epsilon_{xy}}{2}r^2
\right)%+\mathcal{O}\left(\epsilon^2,\epsilon_{xy}^2,\alpha^2\right)\right]
%+\mathcal{O}\left(\left\{\epsilon\right\}^2\right)
\right],
\end{eqnarray}
where $\alpha$ denotes a new  variational parameter and the last expression
is an expansion of the first line to the first order in the deviations.
This ansatz looks somewhat complicated, but this is needed to build in
the relevant physics. This is most easily seen by considering 
the noninteracting limit where the wave functions
are known analytically. 

In the noninteracting limit the vortex states with angular momentum 
projections equal to $\pm N$ are degenerate.
This implies that linear superpositions of these states
have the same energy. As a result, there exists a 
quadrupole-mode
with zero frequency in this limit. To capture this mode the
variational parameter $\alpha$ is included  in the ansatz. To understand this,
assume that $\alpha=0$ and expand the
exponent in Eq.~(\ref{quadrupole_ansatz}). We get
\beq
\exp\left[-\epsilon\frac{\left(x^2-y^2\right)}{2}-\epsilon_{xy}xy\right]
=1-\frac{\epsilon\, r^2}{2}\cos\left(2\phi\right)-\frac{\epsilon_{xy}\, r^2}{2}
\sin\left(2\phi\right)+\mathcal{O}\left(\epsilon^2,
\epsilon_{xy}^2,\epsilon\epsilon_{xy}\right).
\enq
For clarity assume also that
$\epsilon$ is real and $\epsilon_{xy}=-i\epsilon$. The disturbance
then couples to the wave function
\beq
r \exp\left[i\phi\right]\exp\left[-\frac{B_0r^2}{2}\right]
r^2 \exp\left[-2i \phi\right],\nonumber
\enq
which is the wave function of the anti-vortex state multiplied by $r^2$.
This state has obviously a different energy than the true
anti-vortex wave function. As a result the ansatz without $\alpha$ gives
a wrong frequency for this mode in the ideal-gas limit. 
To avoid this problem we need the additional variational parameter
$\alpha$ to give a nonvanishing  amplitude for the correct anti-vortex
wave function in the ideal-gas limit. The fact that this ansatz 
really couples to the correct anti-vortex wave function is most 
clearly seen
by setting $\epsilon=\epsilon_{xy}=0$ in Eq.~(\ref{quadrupole_ansatz}).
Otherwise the ansatz is very similar to the ansatz we used for the
(quasi)condensate without the vortex. In the noninteracting limit the $m=+2$
mode requires coupling to a wave function with angular momentum 
$m=3$ and with a small distance
behaviour that should be proportional to $r^3$. In 
Eq.~(\ref{quadrupole_ansatz}) this is indeed the case, as can be 
verified by setting $\alpha$ equal to $0$.

Using the above ansatz we can determine the quadrupole-modes of a single 
pancake analytically for the full parameter regime 
from the noninteracting limit to the
Thomas-Fermi regime. The equilibrium solution is the same as for the
monopole-mode and is given by Eq.~(\ref{equilibrium_vortex}).
To second order in the deviations 
the various contributions to the Lagrangian are
\begin{eqnarray}
%\beq
%\begin{array}{l}
L_T&=&-2\alpha'\dot{\alpha}''-\frac{3}{2B_0^2}\left(\epsilon'\dot{\epsilon}''
+\epsilon_{xy}'\dot{\epsilon}_{xy}''\right)+\frac{1}{B_0}
\left[\alpha'\left(\dot{\epsilon}''+\dot{\epsilon}_{xy}'\right)+
\alpha''\left(\dot{\epsilon}_{xy}''-\dot{\epsilon}'\right)\right]\nonumber\\
L_V&=&\frac{3}{4B_0^3}\left(|\epsilon|^2+|\epsilon_{xy}|^2\right)
-\frac{1}{2B_0}\left[\alpha'\left(\epsilon'-\epsilon_{xy}''\right)
+\alpha''\left(\epsilon''+\epsilon_{xy}'\right)\right]\nonumber\\
L_K&=&\frac{1}{4B_0}\left(|\epsilon|^2+|\epsilon_{xy}|^2+
4\epsilon_r\epsilon_{xy}''-4\epsilon''\epsilon_{xy}'\right)+
\frac{1}{2}\left[\alpha'\left(\epsilon'-\epsilon_{xy}''\right)+
\alpha''\left(\epsilon''+\epsilon_{xy}'\right)\right]\\
L_{NL}&=&UB_0\left\{|\alpha|^2+\frac{3}{8B_0^2}
\left(\epsilon'^2-\epsilon''^2+\epsilon_{xy}'^2-\epsilon_{xy}''^2\right)
%\right.\\
%\left.\;\;\;\;\;\;\;\;
-\frac{1}{4B_0}\left[\alpha'\left(5\epsilon'+\epsilon_{xy}''\right)
+\alpha''\left(5\epsilon_{xy}'-\epsilon''\right)
\right]\right\},\nonumber
%\end{array}
%\enq  
\end{eqnarray}
where $L_T$ is due to 
the part of the Lagrangian containing the time-derivatives,
$L_V$ is due to 
the potential energy, $L_K$ is due to the kinetic energy, and
$L_{NL}$ is the contribution due to the interactions between atoms.

\begin{figure}
\includegraphics[width=\columnwidth]{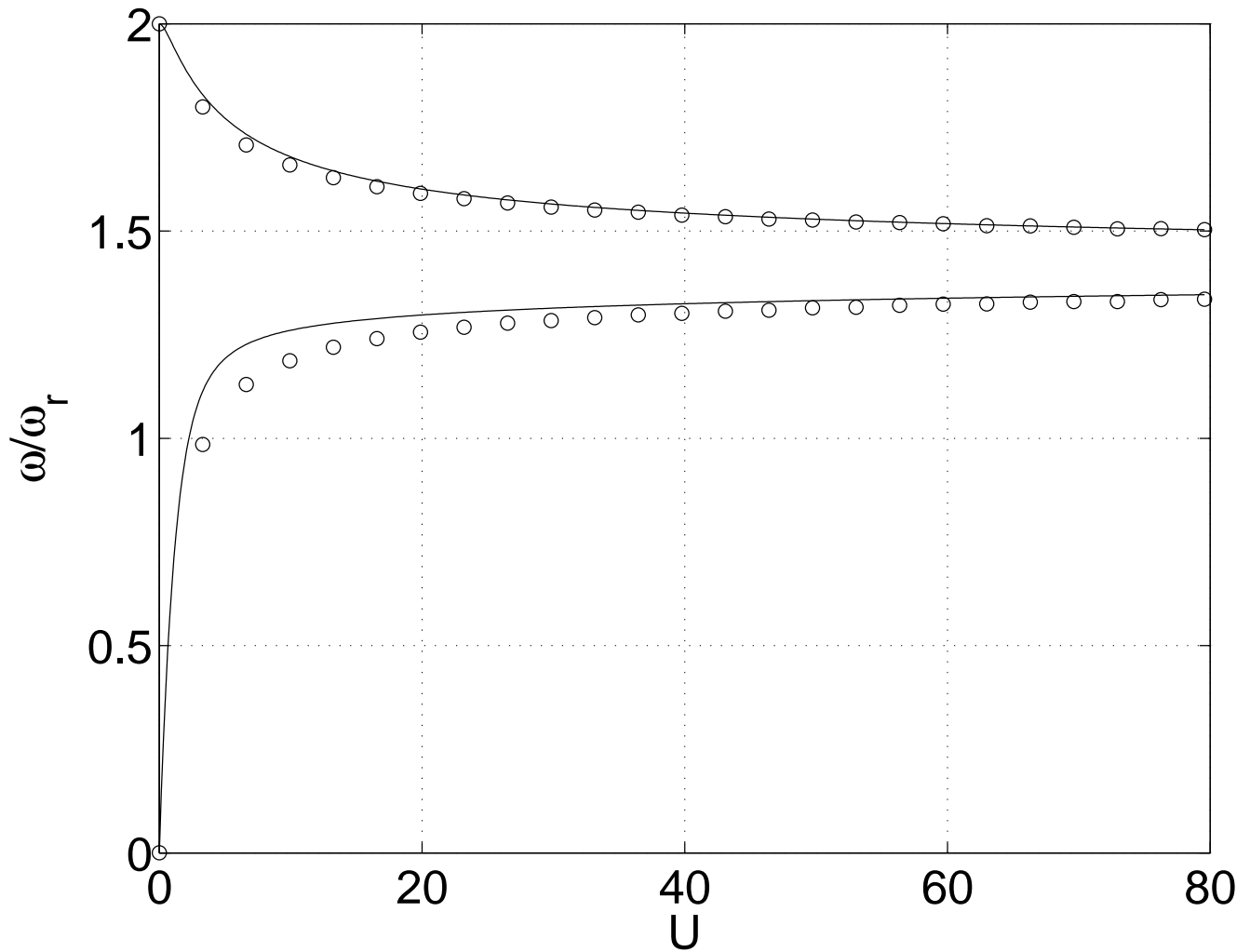}
%\includegraphics[width=10.0cm]{splitting.eps}
%\vspace*{1cm}
\caption[]{Splitting of the quadrupole-modes for the Bose-Einstein 
condensate with a vortex as a function of interaction strength. The solid
line is the analytical result based on the ansatz in 
Eq.~(\ref{quadrupole_ansatz}) and the open circles are calculated 
by solving the Bogoliubov-deGennes equations numerically.
\label{fig:quadrupole_splitting}}
\end{figure}

With this result we can solve for the eigenmodes of the system. The problem 
is essentially that of solving the eigenvalues of a $3\times 3$ matrix.
This matrix has three (generally) nondegenerate eigenvalues and two of 
these correspond to the quadrupole-modes. The third mode is of no
interest to us here. In this modes the deviation from the equilibrium
is a superposition of various trap states, among which 
the $m=5$ component has an incorrect short distance behaviour that causes 
the energy of this mode to strongly increase with increasing atom
number. The frequencies of the quadrupole-modes
can be calculated analytically, but the results are too long to be given here.
However, they do not cause any computational problems.
In Fig.~\ref{fig:quadrupole_splitting} we show the frequencies of
the quadrupole-modes based on our ansatz and compare them against
the values computed numerically with the Bogoliubov-deGennes 
equations~\cite{Dodd1997a}.
The agreement is very good over the whole range of interaction strengths.

In the limit of a nearly ideal-gas the quadrupole frequencies are 
$\omega_{-2}=U$ and $\omega_{2}=2$.
For large atom numbers the quadrupole-mode frequencies are given by
\beq
\label{TF_quadrupole_vortex}
\omega_{\pm 2}=\sqrt{2}\pm \frac{1}{\sqrt{2U}}
\enq
and the splitting between the modes is $\omega_2-\omega_{-2}=\sqrt{2/U}$.
Zambelli and Stringari~\cite{Zambelli1998a} used 
sum rules to show that
the splitting between the quadrupole-modes in the limit of large
atom numbers should be
\beq
\omega_2-\omega_{-2}=\frac{2\langle L_z\rangle}{\langle r^2\rangle}.
\enq
Here $L_z$ is the $z$-component of the angular momentum operator.
With our ansatz we have $\langle L_z\rangle=1$ and $\langle r^2\rangle=2/B_0$
and the splitting of the quadrupole-modes 
is indeed the same as the result based on the sum rule approach.

\subsection{Influence of the lattice on the 
quadrupole-modes of the vortex state}
\label{sec:vortexquadrupolelattice}
\begin{figure}
\includegraphics[width=\columnwidth]{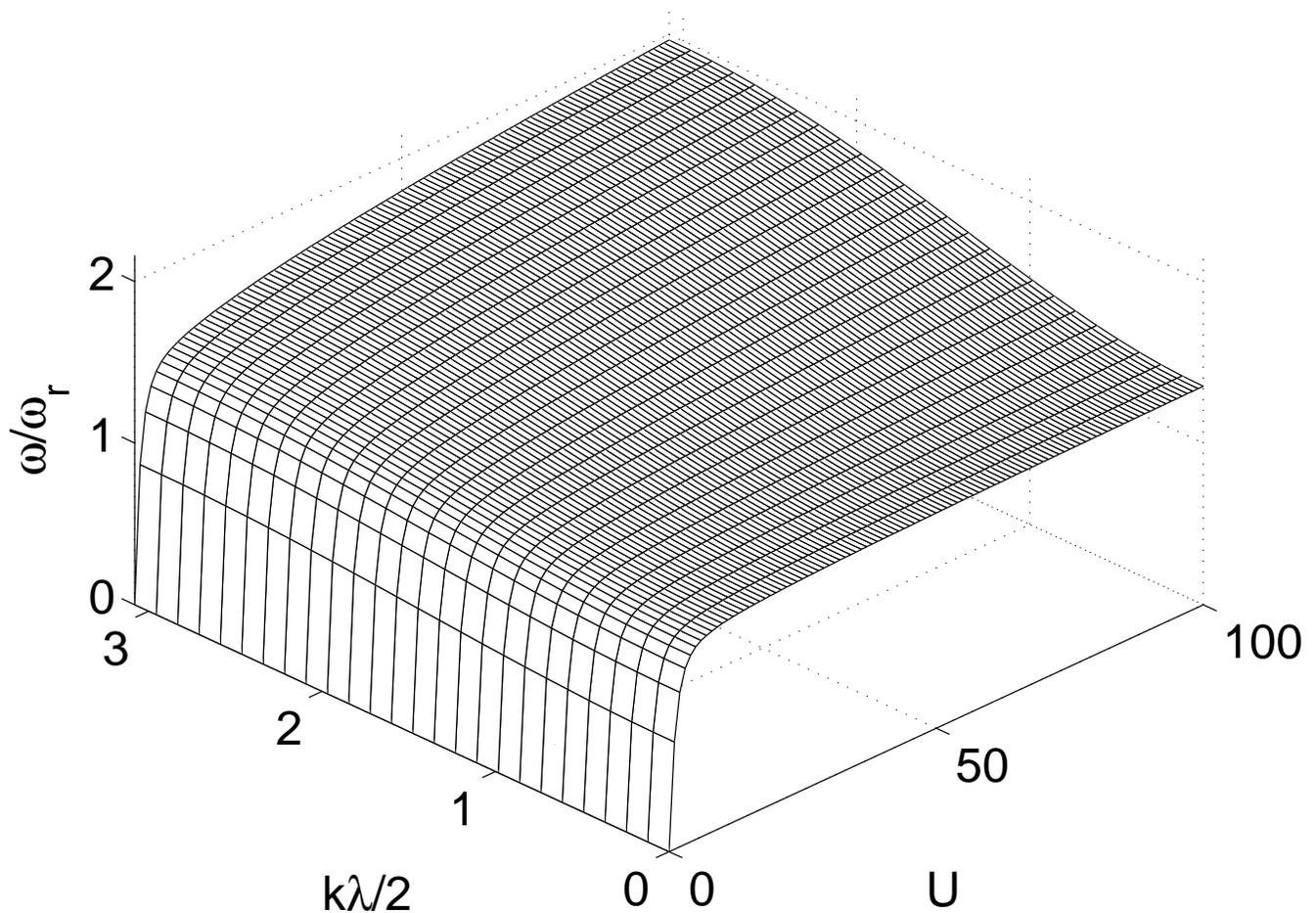}
%\vspace*{1cm}
\caption[]{Dispersion of the quadrupole-mode with $m=-2$ for 
the Bose-Einstein condensate with a vortex when $J=0.05$. 
The figure is based on 
the wave function ansatz in Eq.~(\ref{quadrupole_ansatz}).
\label{fig:quadrupole_lattice_mm2}}
\end{figure}

\begin{figure}
\includegraphics[width=\columnwidth]{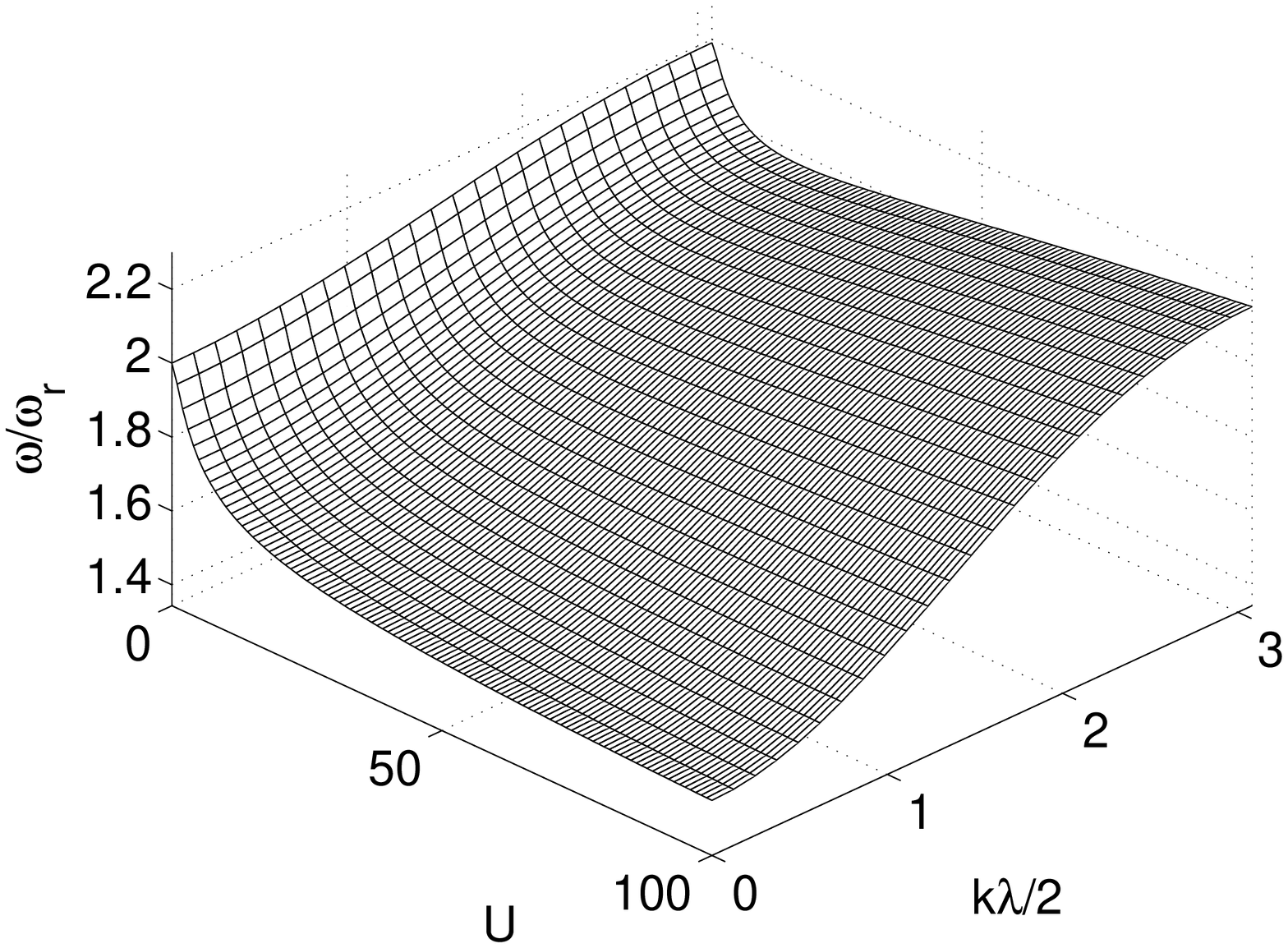}
%\vspace*{1cm}
\caption[]{Dispersion of the quadrupole-mode with $m=2$ for 
the Bose-Einstein condensate with a vortex when $J=0.05$. 
The figure is based on 
the wave function ansatz in Eq.~(\ref{quadrupole_ansatz}). Note that
for clarity the viewing angle is different from the previous figures.
\label{fig:quadrupole_lattice_m2}}
\end{figure}

Including the lattice structure makes the already complicated equations
even more complicated~\cite{comment2003b}. 
In Fourier space the nearest-neighbor interaction
introduces a new term to the Hamiltonian
\begin{eqnarray}
H_{J}&=&-J\sum_k
\left(\cos\left(\frac{k\lambda}{2}\right)-1\right)\left\{
2|\alpha_k|^2+\frac{3}{2B_0^2}\left(|\epsilon_k|^2+|\epsilon_{xy,k}|^2\right)+
\right.\nonumber\\
&+&\left.\frac{2}{B_0}\left[\alpha_{k}'\left(\epsilon_{xy,k}''-
\epsilon_{k}'\right)
-\alpha_{k}''\left(\epsilon_{k}''+\epsilon_{xy,k}'\right)\right]
\right\},
\end{eqnarray}
where $\alpha_{k}'$ is the Fourier transform of the real part of $\alpha_n$,
$\epsilon_{k}'$ is the Fourier-transform of the real part of $\epsilon$,
and $\epsilon_{xy,k}'$ 
is the Fourier-transform of the real part of $\epsilon_{xy}$.
Similar notation applies to the imaginary parts and for example 
$|\epsilon_k|^2=\epsilon_{k}'\epsilon_{-k}'+\epsilon_{k}''\epsilon_{-k}''$.
In Figs~ \ref{fig:quadrupole_lattice_mm2} and 
~\ref{fig:quadrupole_lattice_m2} we show  in detail the resulting dispersion
relations for the quadrupole-modes  
as a function of $k$ and interaction strength $U$. 

%Two of the eigenvalues of the matrix 
%\beq
%O^2=-\left(\begin{array}{ccc}
%\frac{2}{B_{0}^{2}}\left[1-4B_{0}^{2}+B_{0}^{4}
%-J\left(3B_{0}^{2}+B_{0}\right)C_f\right] & \frac{2J}{B_0}
%\left(1-B_0^2\right)C_f  & 
%\frac{4}{B_0}\left(B_0^2-1\right)\left[B_0^2-1+
%JB_0C_f\right]
%\\
%\frac{B_0^2-1}{B_0^2}\left(1+B_0^2-2JB_0C_f\right) & 
%\frac{-2}{B_0}\left[B_0+B_0^3-J\left(3B_0^2+1\right)C_f\right] & 
%\frac{2}{B_0}\left(B_0^2-1\right)\left(1+B_0^2-2JB_0C_f\right)
%\\
%\frac{1}{B_0^3}\left[3-9B_0^2+4B_0^4-JB_0\left(3-7B_0^2\right)C_f\right]
%& \frac{1}{B_0^2}\left[2B_0^3+J\left(3-7B_0^2\right)C_f\right]
%& \frac{2}{B_0^2}\left(1-B_0^2\right)\left(-3+2B_0^2+6JB_0C_f\right)
%\end{array}\right)
%\enq
%correspond to the quadrupole-modes. We used a shorthand
%\beq
%C_f=\cos\left(\frac{k\lambda}{2}\right)-1.
%\enq

Even though the general formulae are too complictated to be given here,
the ideal-gas limit and the Thomas-Fermi limit give us simple formulae.
In the limit of weak interactions we have 
$\omega_{-2}=U-2J\left(\cos\left(k\lambda/2\right)-1\right)$
and $\omega_{2}=2-2J\left(\cos\left(k\lambda/2\right)-1\right)$ and
in the limit of strong interactions or large particle numbers we have
\beq
\omega_{\pm 2,v}=\sqrt{2}\left[1-\frac{3J}{4}\sqrt{\frac{U}{2}}
\left(\cos\left(\frac{k\lambda}{2}\right)-1\right)
\right].
\enq
In the limit of large particle numbers the effective mass of the
quadrupole-modes thus becomes
\beq
m_{\pm 2,v}^*=\frac{16}{3J\sqrt{U}}
\left(\frac{\hbar}{\omega_r\lambda^2}\right).
\enq
This result indicates that we expect the quadrupole-modes to have
about three times larger effective mass than the breathing mode.
Again this can be understood by overlap arguments.


%\subsection{Influence of the lattice on the excitation frequencies}
%\label{sec:vortexlattice}

\section{Summary and conclusions}
\label{sec:conclusions}
We have calculated the band structure of the most important
transverse collective excitations of a stack of two-dimensional 
Bose-Einstein condensates
in a one-dimensional optical lattice with and without a vortex. 
Our variational approach enables us to crossover smoothly from the ideal-gas
into the Thomas-Fermi regime and to treat the interlayer
coupling without other approximations that those involved in the
variational ansatz. We have also calculated the 
short wave length part of the excitation spectra, which means that
in our approach neighboring sites can be completely out of phase
with each other.
Using our general results for the excitation frequencies, 
we derived predictions for the
the effective mass of the monopole and quadrupole-modes. We 
noticed that the effective mass is sensitive to the mode in question
as well as to the presence of a vortex.
In this paper we have only focused on the linear response of the
system. For large modulations nonlinear effects can become 
important~\cite{Wu2001a,Konotop2002a,Baizakov2002a}. In particular,
assumptions about a nearly homogeneous condensate can break down as
the system becomes dynamically unstable towards large density
modulations.

Experimentally the kind of excitations we have discussed in this paper
can be created by modulating the radial trapping frequency 
$\omega_r$ as a function of $z$. One possible way to excite the monopole 
modes is to have two counter-propagating laser beams with a Gaussian
intensity profile. Due to the optical dipole force 
the intensity profile of each one of the beams would 
provide the trapping in the radial direction, while the interference between
the beams would provide the necessary modulation. To excite
equal superposition of $m=\pm 2$ quadrupole-modes sheets of laser-light
can be considered.

\begin{figure}
\includegraphics[width=\columnwidth]{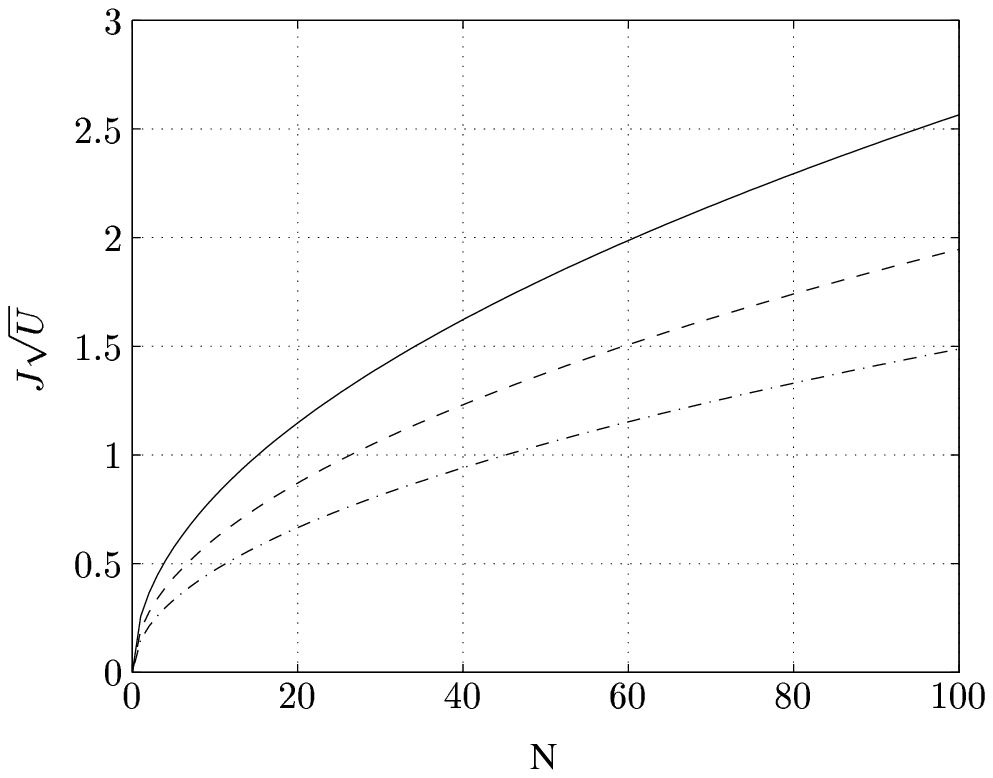}
%\vspace*{1cm}
\caption[]{The quantity $J\sqrt{U}$ as a function of the number of 
$^{87}{\rm Rb}$ atoms
in the each lattice site for 
different lattice depth. The solid line is for the depth 
$V_0=8 E_r$, the dashed line is for $V_0=9 E_r$, and the dot dashed line
is for $V_0=10 E_r$. The wave length of the laser-light was taken to be
$\lambda=800{\rm nm}$ and the radial trapping frequency was 
$\omega_r/2\pi=100 {\rm Hz}$.
\label{fig:JsqrtU}}
\end{figure}
In the limit of large interactions the constant in front of the
$k$ dependent part of the dispersion relations 
always scales with $J\sqrt{U}$.
This number is a good measure of how strong the effects due to the
lattice are. If this number is small, the lattice 
effects are hard to distinguish
experimentally from the dominant single site result.
In Fig.~\ref{fig:JsqrtU} we plot $J\sqrt{U}$ as a function of 
the on-site atom number for a few different lattice depths. As can be
seen, the effects of the lattice for the modes we are considering
can be very pronounced and should be easily observable.

%In the paper we have ignored fluctuations of the on-site atom numbers.
%Fluctuations of the on-site atom numbers are accompanied by
%superfluid flows along the lattice. If such flows are too
%rapid one might encounter dynamical 
%instabilities~\cite{Wu2001a,Konotop2002a,Baizakov2002a}.
%To avoid such effects the modulation the condensate
%has to be small. In our model the phase difference between neighboring sites
%is largest at the edge of the pancakes. This gives us a rough
%criteria $\Delta \phi\sim \epsilon_i/B_0\ll 1$. For example,
%assume exciting a breathing mode where the size of the
%condensate is initially changed by a factor $\Delta R/R_{TF}$ and then the 
%pancake is allowed to evolve freely. To avoid dynamical instabilities
%one should have $\left(\Delta R/R_{TF}\right)/B_0\ll 1$. 

In the system we have discussed in this paper, each 
(quasi)condensate becomes very
quickly two-dimensional as the depth of the lattice is increased.
In particular the coherence length in the center of the 
two-dimensional (quasi)condensate quickly 
becomes larger than the thickness of the pancake. In low
dimensions phase fluctuations are expected to be more 
pronounced~\cite{Andersen2002a,Khawaja2002a,Khawaja2003a}. 
In our treatment we ignore such fluctuations. In a two-dimensional system
there is a true condensate at zero temperature and then the phase fluctuations
are not expected to play a major role. At nonzero temperatures
phase fluctuations become more important, but are expected
to be more pronounced between sites that are well separated.
In our parameter regime the tunneling term coupling neighboring sites
will establish phase coherence between neighbors. As a result the 
two-dimensional (quasi)condensates
are not strictly two-dimensional since they ``see'' the third direction
through the tunneling term.
As the distance between the sites increases the 
phases become less correlated, but
as we are only interested in the nearest-neighbor couplings, such
effects are not important. Consequently, we expect our model to
be applicable also at small but nonzero temperatures. Phase fluctuations
may cause a slight reduction 
in the strength of the Josephson coupling, but would
leave our results otherwise unchanged.

In this paper we have choosen to fix the number of atoms in 
every two-dimensional (quasi)condensate. In our variational approach it is not
difficult to include atom number as well as global phase fluctuations
by replacing in our variational ansatz $\sqrt{N}$ by 
$\sqrt{N_n(t)}e^{i\nu_n(t)}$, where $N_n$ denotes the number of atoms 
and $\nu_n$ the global phase of the (quasi)condensate in every site.
In the simplest case where we neglect the couplings with the transverse
modes, we find that at long wave lengths there exists a 
phonon mode with the sound velocity
\beq
c_s=\sqrt{\frac{UJ}{\sqrt{1+2U}}
\left(\frac{\hbar\omega_r}{m}\right)}\;\lambda/l_r,
\enq
which agrees exactly with the results
obtained previously~\cite{Javanainen1999a,vanOosten2001a,Kraemer2002a}.

In a  recent experiment the 
Kelvin modes of a Bose-Einstein condensate with a vortex were
observed~\cite{Bretin2003a}.
In the model that we have presented in this paper 
the vortex is always in the center of
each pancake. In the future we plan to relax this condition
and consider also the dynamics of the vortex. In this manner it
is possible to
study the Kelvin modes in an optical lattice, and in particular
their coupling to the transverse excitations which were our main
focus here.

\begin{acknowledgments}
This work is supported by the Stichting voor Fundamenteel Onderzoek der 
Materie (FOM), which is supported by the Nederlandse Organisatie voor 
Wetenschaplijk Onderzoek (NWO).
\end{acknowledgments}

%\appendix

%\bibliographystyle{apsrev}
%\bibliography{bibli}

\begin{thebibliography}{38}
\bibitem[{com({\natexlab{a}})}]{comment2003a}
\bibinfo{note}{In addition, the chemical potential must be smaller than the
  energy splitting $\hbar\omega_L$. With typical parameters this condition is
  always met as long as the chemical potential is much smaller than the lattice
  depth.}
\end{thebibliography}

\end{document}